# Establishment of Imaging Spectroscopy of Nuclear Gamma-Rays based on Geometrical Optics

*Toru Tanimori[1,2], Yoshitaka Mizumura[2,1], Atsushi Takada[1], Shohei Miyamoto[1], Taito Takemura[1], Tetsuro Kishimoto[1], Shotaro Komura[1], Hidetoshi Kubo[1], Shunsuke Kurosawa[3], Yoshihiro Matsuoka[1], Kentaro Miuchi[4], Tetsuya Mizumoto[1], Yuma Nakamasu[1], Kiseki Nakamura[1], Joseph D. Parker[1], Tatsuya Sawano[5], Shinya Sonoda[1], Dai Tomono[1], Kei Yoshikawa[1]

1 Graduate School of Science, Kyoto University, Sakyo, Kyoto 606-8502, Japan

2 Unit of Synergetic Studies for Space, Kyoto University, Sakyo, Kyoto, 606-8502, Japan

3 Institute of Materials Research, Tohoku University, Sendai, Miyagi, 980-8577, Japan

4 Department of Physics, Kobe University, Kobe, Hyogo, 658-8501, Japan

5 Department of Physics, Kanazawa University, Kanazawa, Ishikawa, 920-1192, Japan

Since the discovery of nuclear gamma-rays, its imaging has been limited to pseudo imaging, such as Compton Camera (CC) and coded mask. Pseudo imaging does not keep physical information (intensity, or brightness in Optics) along a ray, and thus is capable of no more than qualitative imaging of bright objects. To attain quantitative imaging, cameras that realize geometrical optics is essential, which would be, for nuclear MeV gammas, only possible via complete reconstruction of the Compton process. Recently we have revealed that "Electron Tracking Compton Camera" (ETCC) provides a well-defined Point Spread Function (PSF). The information of an incoming gamma is kept along a ray with the PSF and that is equivalent to geometrical optics. Here we present an imaging-spectroscopic measurement with the ETCC. Our results highlight the intrinsic difficulty with CCs in performing accurate imaging, and show that the ETCC surmounts this problem. The imaging capability also helps the ETCC suppress the noise level dramatically by ~3 orders of magnitude without a shielding structure. Furthermore, full reconstruction of Compton process with the ETCC provides spectra free of Compton edges. These results mark the first proper imaging of nuclear gammas based on the genuine geometrical optics.

**Introduction**

Nuclear gamma-rays were discovered in 1890s, and since then many scientists have made a great effort to invent a fine imaging method of nuclear gammas with little success. For nuclear gammas, imaging methods have been still limited to pseudo imaging, such as collimators and coded masks, which are capable of no more than qualitative imaging of bright and point-like objects. Compton Camera (CC), of which the basic concept was proposed in 1974, is an advanced gamma imager[1] based on partially-geometrical optics and has a potential of a breakthrough. Nevertheless even the basic requirement of quantitative evaluation of the gamma intensity in an image has not been achieved with the CC so far despite the fact that many improvements have been made[2-6] over decades.

Proper imaging is defined as a mapping of rays with different incident angles defined by two angles (polar angle $\zeta$ and azimuthal one $\eta$ in Fig.1a) to separate unique points on a plane at infinity ("imaging plane" or "plane of hemisphere" in astronomy), as is trivially the case for optical telescopes. Commonly, a photon from radio to X-rays is mapped to a single unique point, corresponding to the incident angle, on the imaging plane by reflectors or lenses. In gamma-rays the situation is very different; an incident photon is usually split to multiple rays and/or particles in a detector. Therefore, proper imaging of gamma-rays is far from trivial. A MeV gamma interacts with a matter via the Compton process and transforms into a recoil electron and a Compton-scattered gamma. In handling the kinematics of Compton scattering, the de facto standard coordinate system is defined event by event along a plane made by the two directions of incident and scattered gammas (hereafter dubbed the Compton coordinates; see Fig.1b). Its relative position varies event by event with respect to the absolute coordinate system of the imaging plane which is fixed to the detector (hereafter, the laboratory coordinates). Since the two directional angles ($\zeta$ and $\eta$) of the incident gamma are defined on the imaging plane (Fig.1a), they must be calculated from the measured parameters of Compton-scattering polar angle $\theta$ and electron-azimuthal angle $\phi$ defined on the Compton coordinates event by event according to the equation explained in Section Method. This complex interaction of a gamma with a matter makes proper imaging of MeV gammas very challenging. In principle, it can be achieved by resolving the equation of the Compton process event by event, where both the directions and energies for a pair of a recoil electron and Compton-scattered gamma for each incoming gamma are required to be measured. However, tracks of electrons are hardly measured with existing instruments. Even cutting-edge Compton Cameras (CCs) measure only $\zeta$ corresponding to the Compton-scattering angle ($\theta$) in the Compton process, as illustrated in Figs.1a, 1b and 1c. It is in stark contrast to imaging of GeV gammas. Proper imaging has been long achieved for GeV gammas by measuring the two angles of incident gammas, tracking both the electron and position after pair-creation. Then, GeV gammas are electrically focused on the imaging plane by reconstructing the pair-creation process. The method has been used in high-energy gamma-ray astronomy; indeed the latest astronomical GeV-gamma space observatory "Fermi"[7] is equipped with large multi-layer silicon strip detectors (SSD) for tracking electrons and positrons, which provide the PSF with the size of <1°, and Fermi has successfully detected several thousand objects.

The first major application of the CC was for MeV gamma-ray astronomical observations by COMPTEL onboard Compton Gamma-ray Observatory satellite launched in 1991[8]. In the early phase of the project, optimization algorithms, such as Maximum Entropy method (MEM), were adopted in order to compensate the lack of information of the azimuthal angle (η in Fig.1a and 1c) of the incident gammas. They successfully detected three sources in a pre-launch laboratory experiment, where the background was ~3 times as intense as the sources, by adopting the MEM. However, the use of the MEM turned out to be limited and be only for obtaining the qualitative image at best during their observation programme[9-12]. Eventually they came to use only the likelihood method for any quantitative results, which took count of the effects of the spread of Compton scattering angle θ and extended backgrounds.

In celestial MeV gamma-ray observations, the background is usually stronger by two orders of magnitude or more than actual sources, and in these circumstances no optimization algorithms, including the MEM and maximum likelihood expectation maximization (MLEM), work well, as discussed in detail in Section Results. On the other hand, the simple likelihood method, which optimizes the probabilities of the signal and the background within the allowable statistical limit, seems to be most reasonable and reliable although the detail of how they applied it in the astronomical observations of COMPTEL has not been published. After COMPTEL, hardly any sound results have been published about applications of MLEM to CCs, except for some qualitative and visual improvement on images. Despite the lack of any quantitative proofs, the unfound expectation seems to be popular among the scientists in the field that an optimization algorithm, such as MLEM, would eventually enable us to achieve the PSF of CCs with the angular resolution of the polar angle ζ, which is the same as the resolution of the Compton Scattering angle θ (generally called Angular Resolution Measure: ARM)[8]. Indeed, the recent proposal for a satellite of MeV gamma-ray astronomy with CCs seems to treat the ARM in the same way as the PSF in the geometrical optics.

In X-ray astronomy, imaging spectroscopy was realized, using reflectors, which focuses X-ray photons onto the focal point. The use of reflectors improves the sensitivity by an order of two or more, compared with those of multi pin-hole and coded-mask detectors. A pin-hole camera is intrinsically capable of imaging spectroscopy, keeping the directional information of each incoming photon on the imaging plane. The tradeoff is, however, a seriously small effective area and hence low sensitivity. In coded mask detectors, on the other hand, different directional photons passing through different holes can be projected onto the same point on the imaging plane. This breaks the conservation of the intensity and hence makes imaging spectroscopy difficult. In CCs, many different directional circles overlap at the same point on the imaging plane, and hence it does not conserve the intensity (Figs.1a, 1d and 4b). As such, present nuclear imaging methods except pin-hole cameras mix up the information of different directional gammas at a point on the imaging plane, and thus, the spectroscopic features can be measured only for the whole field of view.

In order to attain proper imaging of nuclear gammas based on genuine geometrical optics, we have developed Electron-Tracking Compton Camera (ETCC) (Tanimori, T. *et al*. 2015[13], hereafter referred to as TT), which outputs the two angles of an incident gamma, ζ and η, by measuring the direction of a recoil electron, and accordingly provides the brightness distribution of gammas with a resolution of the PSF. The most distinctive feature of the ETCC is to resolve the Compton process fully by measuring a 3-D recoil electron track in the gas. Here we report our successful nuclear gamma-ray spectroscopy with the ETCC in the way similar to that with optical or X-ray telescopes.

**Instrument**

Measurement of the azimuthal angle η (Fig.1a) is crucial to obtain a fine and quantitative image in the nuclear gamma-ray band. It can be derived from the momentum vector of a recoil electron (ϕ in Fig.1b) and the Compton scattering point (see the section Method for detail). Also, the reduction of the background is required with a relevant effective area of a few tens of $cm^2$ similar to that of a crystal or Ge detector. The ambiguity of ϕ is generally called the Scatter Plane Deviation (SPD), which extends along the circle of the Compton annulus (Fig.1b).

In 1990s the TIGRE group discussed the advantage of the measurement of a recoil electron to reduce the background, based on the simulation study[14], while developing an advanced CC consisting of multi-layer silicon strip detectors (SSD). Then, the MEGA group measured the tracks of recoil electrons with the energy higher than 2 MeV, using a similar CC consisting of multi-layer SSD and CsI(Tl) calorimeter until 2003[15]. They found that a few sampling points of the track caused serious confusions in resolving a correct permutation of those hit points along the track[16]. These confusions then increase the number of possible permutations of the track and the scattering points in each event, and make the ARM resolution deteriorate due to the uncertainty in the scattering point. Thus, a tracking detector which can measure a very fine tracking even for very low energy recoil electrons of ~10 keV, such as a cloud chamber, is needed to attain a perfect electron tracking and then to reconstruct almost all the incoming nuclear gammas in the sub-MeV and MeV energy regions.

The ETCC is a unique instrument to satisfy these requirements. Here we describe its principle and structure briefly. The ETCC has similar components to CCs, namely a forward detector as a scatterer of nuclear gammas and a backward detector, which works as a calorimeter to measure the energy and hit position of each scattered gamma. Our ETCC comprises, as a forward detector, a gaseous Time Projection Chamber (TPC), which is based on micro-pattern gas detectors (MPGD), in order to measure 3-D tracks of recoil electrons, and as a backward detector, pixel scintillator arrays (PSAs) with heavy crystal (at present we use $Gd_2SiO_5$:Ce, GSO). Until now we have used Ar based gas for TPCs used in all ETCCs mentioned later, and its energy resolution is ~30% at 5.9 keV. The energy resolution of the ETCC with GSO scintillators[13] is 11% in full-width half maximum (FWHM) at 662 keV.

In 2004 we reported the first successful full electron-tracking in the laboratory experiment with a prototype 10-cm-cubic ETCC. We then carried out "Sub-MeV gamma ray Imaging Loaded-on-balloon Experiment" with the improved 10-cm-cubic ETCC (SMILE-I system) to measure the diffuse cosmic MeV gamma-rays with a balloon-borne experiment in 2006[17], and demonstrated its excellent ability of particle identification, based on dE/dx of an electron track in the gas, where the background level was dramatically reduced by an order of ~3 without an active shield. Although its detection efficiency in the sub-MeV band was lower than $10^{-4}$ due to the low detection efficiency of an electron track in the TPC of SMILE-I, different types of gas could in principle give a higher efficiency (larger effective area); for example, the effective areas of 110 cm$^2$ and 65 cm$^2$ are expected at 1 MeV for a 50-cm-cubic ETCC with 3-atm $CF_4$ and Ar gases, respectively. $CF_4$ has a 2.3 times larger cross section of Compton scattering than Ar gas.

We then developed a 30-cm-cubic ETCC (Fig. 1 in TT) to achieve an effective area of a few cm$^2$ at 300 keV, aiming to detect celestial MeV-gamma-ray objects such as the Crab, with a balloon experiment (SMILE-II)[13, 18]. All the data presented in this paper are based on this 30-cm-cubic ETCC. In this ETCC, tracking efficiency in the TPC is improved drastically to 100% from 10% in SMILE-I, owing to the improved readout electronics and algorithm[13, 18], and accordingly the ETCC provides a better noise reduction of dE/dx and good angular resolutions of ARM of 5.3º (FWHM) at 662 keV, which is broadly consistent with that calculated from the detector energy resolution.

The next step is to improve the SPD, because a better SPD is more critical than a better ARM to attain a better PSF (see TT). In the MPGD of the TPC, orthogonal strips of the X and Y coordinates are used in order to reduce the number of readout channels in the MPGD. A problem is that this readout method causes a well-known uncertainty in tracking due to multi-hits on the same timing, which increased the SPD considerably to 200º (FWHM) with SMILE-I. In the new readout electronics of SMILE-II, a pulse width of each electrode in the MPGD is recorded as a timing width over the threshold (TOT), which provides a coarse charge deposit at each hit point. Using TOT, we have reduced the coincidence timing width between the X and Y strips from 10 ns to ~1 ns, and have improved the SPD to ~100º (FWHM)[13], and accordingly have reduced the uncertainty in tracking.

Consequently, we have obtained an effective area of ~1 cm$^2$ at 300 keV and the half-power radius of 15º for the PSF. With this sensitivity, gamma-rays from the Crab is estimated to be detected at a significance of ~5σ with a several-hours-long balloon observation[13]. This unprecedentedly high-level performance of the ETCC lead us to our "discovery" of the PSF in the same sense as in optical telescopes, which no one had ever seriously thought of in MeV-gamma cameras. It is a significant leap in technological development in the field of nuclear gamma-ray spectroscopy, and is the focal point discussed in detail in the following sections.

**Results**

Since CCs measure only a polar angle ζ from the measured Compton-scattering angle θ, the other direction of a gamma (η) is obtained by accumulating the Compton annuli as a cross point of those circles, of which the radius and the width are θ and its angular resolution ARM, respectively. Thus, CC attains only partial imaging of the projection of the gamma direction to a circle, rather than a point. Then we have to use the annulus as a probability distribution function to conserve the number of events, which means that any event distribution on the imaging plane measured by CCs is highly smeared by the wide angle of θ. This is a reason why images measured by CCs appear to be much smoother than the statistical fluctuation determined by the number of events. On the other hand, ETCCs obtain the direction of a gamma as a point on the imaging plane with ζ and η. In TT, we used a probability distribution of the arc shape since its arc was extended over 200° along the azimuthal direction (SPD) in the early phase of development of the 30-cm-cubic ETCC. However, the ETCC with the much improved SPD with the FWHM of <100° provides a good deal of concentration of gamma events at the focal point on the imaging plane, as shown in Fig.2a and 2c, even when a point direction for the ETCC is used instead of a probability distribution function. Generally, when the probability distribution function on the imaging plane is used to accumulate gammas, the resultant image ends up being spatially smeared doubly with the resolution of the PSF, namely once by the PSF (intrinsic resolution) and once again when the image is smoothed to compensate the poor statistics. In our analysis we create images by simply accumulating directional points of gammas on the imaging plane for the ETCC, and hence avoid over-smearing of images. For CC, we use a line of circle, instead of an annulus, for the direction of a gamma, and remove the smearing effect by the ARM resolution. Then, the difference of the performances between the ETCC and CC is attributed to only the smearing effects for the direction of θ.

With this image accumulation, the PSF(Θ) of the ETCC can be defined clearly; we define it as the angular region that encompasses a half of the gamma events emitted from a distant point source, which is conceptually the same as those in X-ray and GeV gamma telescopes. Therefore, the same concept of imaging spectroscopy as with X-ray telescopes is applied to the nuclear-gamma imaging spectroscopy of the ETCC. For the sake of comparison, we also define the PSF(Θ) of the CC: a circular area that would encompass 50% in probability of gammas emitted from a point source within the angular radius Θ, as illustrated in Figs.1a and 1d.

PSF(Θ) of the ETCC and CC are measured as follows. Let us define δ as the angular distance between the reconstruction direction and the real direction (Fig.1b). Figures 2a and 2b are the scatter plots of δ versus the Compton-scattering angle θ for gammas from $^{137}$Cs (662 keV), measured by a 30-cm-cubic ETCC with ARM of 5° (FWHM), SPD of 100° (FWHM), and Θ ~15° at 662 keV[13]. We define the imaging analysis using two angles (θ, φ) of the gamma direction as "ETCC analysis" and that using only one angle θ as "CC analysis". The difference between the two analyses will highlight the power of use of electron-tracking information. The measured δ in the CC analysis is found to spread over the Field of View (FoV), while that in the ETCC analysis concentrates at the centre (Fig.2c). The mean of δ corresponds to the half-power radius Θ of the PSF. The correlation δ=2θ is apparent in the CC for any angles of θ, and therefore the PSF of the CC depends on the Compton-scattering angle θ. In contrast, the parameter δ in the ETCC stays within the size of the PSF independently of θ. Thus, the PSF of the ETCC is uniquely determined from the ARM and SPD (Fig.2a) . Figure 2d shows the measured variations of Θ for the CC and ETCC as a function of gamma energy with the simulation results derived with hypothetically improved ETCC (2° of ARM and 15° of SPD at 662 keV). The simulation program that we used is based on GEANT4[19], and was developed for the SMILE-I and II projects. Its performance has been verified in Takada, A. *et al*. 2011[17], Mizumoto, T. *et al*., 2015[20], and TT. No significant improvement is found in the PSF in the CC analysis over those obtained with conventional CCs, presumably because the PSF of the CC depends on the Compton-scattering angle θ, and always spreads up to ~30°, even though conventional CCs usually have a better ARM than the ETCC.

In CCs, many event circles overlap, and that makes the PSF extended. Therefore, it is difficult to estimate the sensitivity from its effective area and the ARM. Some teams have applied optimization algorithms like MEM and MLEM to solve the complicated overlaps[8, 21], but have found that no algorithms intrinsically improve the statistics. For the detector with an effective area of $A_{eff}$ and PSF(Θ), the significance of signal is proportional to $F_g \cdot \sqrt{A_{eff}}/\sqrt{(F_g+B_g \cdot \Theta^2)}$, where $F_g$ is the signal flux and $B_g$ is the background intensity. For signal-dominated and background-dominated cases, this term is approximated to $\sqrt{(A_{eff} \cdot F_g)}$, and $F_g \cdot \sqrt{A_{eff}}/(\Theta \cdot \sqrt{B_g})$, respectively. In the former case, the significance does not depend on Θ, and the simulation (Fig.3a) suggests that the optimization algorithms seem to be functional. In the latter case, the significance explicitly depends on Θ. In fact, background events can pile up anywhere in the CC (see the next paragraph and Figs.4a and b). Therefore, when we simply count the number of gammas of which the circle has passed through the source region with the radius of ARM, the number of background events should also increase roughly by a factor of $(2\pi\theta \times ARM)/\pi\left(\frac{ARM}{2}\right)^2$ ~100 (Compton-scattering angle θ=40°, ARM=2°) for a background-dominated observation. Thus, the increase of the background very strongly affects the sensitivity and significance of detection (roughly by a factor of 10). We note that the traditional approach for this is to integrate the probability that gammas cover the region (see the previous section), which has been adopted also in recent proposals of advanced CCs. Figure 3b is the same as Fig.3a but with the increased background levels of $10^2$ and $10^3$ of the source with a simulation, in which the significances with the CC analysis are found to deteriorate dramatically.

This difference of the signal significance between the CC and ETCC is explained well with respect to the difference in their spectroscopic responses, as follows. The ETCC is an electrically-focusing telescope of gammas into the PSF, whereas gammas observed by a CC spread far beyond the source. Figure 4a shows a gamma-ray image of $^{137}$Cs (662 keV) placed at a distance of 2 m. We derived the energy spectra each within a 15°-radius circle with the centre at various angular distances from 0° (source position) to 60°. Fig.4c and 4d show the obtained spectra for three angular distances, derived with the ETCC and CC analyses, respectively. A peak at 662 keV was found to diminish sharply in the ETCC analysis as the angular distance of the region increases, but was found not to change much in the CC analysis. Furthermore, low energy gammas, which are scattered gammas in the air, do not decrease in any angular distances in the CC analysis in spite of the low background level in our experiment. When the same energy density of the background as that of the source at the centre in Fig.4a is distributed uniformly up to 80°, the fake peak in the spectrum made by the background at the centre (source position) becomes 3 times larger than the true peak made by the source in the CC, whereas that in the ETCC with $\Theta=15°$ increases by only ~15%. Whereas a better ARM hardly improves it for the CC, the smaller PSF radius of $\Theta=5°$ would dramatically reduce it down to ~2%. Note that the signal-to-noise ratio is unity only at the source position in these cases.

These results confirm our interpretation of the CC and ETCC, the latter of which benefits from the superior PSF. Usually backgrounds include gammas with similar energies to those of source gammas, and then some amount of the background spreading over the FoV certainly contaminates the spectrum of the source region. At large angular distances, the energy spectra by the CC are interfered greatly by background gammas, including those coming from the outside of the FoV, and also the accuracy in the measurement of the source intensity deteriorates considerably. It is especially a serious problem in background-dominated cases like gamma-ray astronomy. In contrast, imaging spectroscopy with the ETCC provides the correct energy-spectrum at a region of interest, coupled with efficient background rejection similar to optical and X-ray telescopes. It is an advantage of true geometrical optics for gammas.

Figures 5a and b show an image of a $^{137}$Cs source and its energy spectra for regions at 4 different angular radii, respectively, both extracted with the ETCC. At the radius similar to the PSF (15°), most air-scattered gammas are not contaminated, and a half of 662 keV gammas remains without Compton edges and other lines, which in general appear in conventional gamma spectroscopy. Even the remaining air-scattered gammas can be removed by subtracting events accumulated from the region symmetrical about the centre of the FoV from the source (see Figs.5c and 5d). Thus, the ETCC is a proper-imaging gamma telescope. Compton edges are often observed in the energy spectra taken with CCs. Both the CC and ETCC must capture Compton-scattered gammas in the backward detector, and hence Compton edges should not appear in any of the obtained energy spectra. Compton edges appear only when recoil electrons, instead of Compton-scattered gammas, are measured at the backward detector.

Finally, we discuss the comparison of the ETCC with typical non-imaging spectrometers, which are widely used in the medical field and environmental monitoring of radio-isotopes. In general, imaging spectroscopy reduces backgrounds by a factor proportional to the ratio of the PSF to $4\pi$ ($10^{-2}$ to $10^{-4}$) for distant gamma sources, such as astronomical objects. For nearby sources with the size $A$ and distance $R$ from the detector, an ETCC projects its image by back-projection method on a plane at the distance of R from the detector, where the background is also reduced to $A/(4\pi R^2) \sim 10^{-3}$ of the spectrum taken with a non-imaging spectrometer, for a typical case of $A$ of a few cm$^2$ and $R \sim 15$ cm, while a large portion (a few ten per cent) of the gammas emitted from the source is maintained. Fig.6a demonstrates the power of imaging spectroscopy, presenting the energy spectra, which contain only several emission lines and without Compton edges, of $^{152}$Eu (0.72 MBq) placed at a distance of 2.4 m from the ETCC, before and after the imaging-selection filter is applied. The filter reduces the background by ~2 orders of magnitude, and faint lines above 500 keV are clearly detected. Their line-peak intensities are typically ~$10^{-2}$ of the natural radiation in our laboratory in Kyoto, Japan. Figure 6b shows a spectrum, from which the background has been subtracted, following the same method as in Fig.5d. Further reduction of noise is presented in Fig.7 in TT with a super-faint point source (27 kBq $^{22}$Na), of which the radiation intensity at the ETCC was <10$^{-3}$ of the typical natural radiation.

Energy resolutions of Ge-semiconductor and high-quality scintillators are a few keV and a few 10 keV, respectively, for MeV gammas. The effective area of a 30-cm-cubic ETCC with 3-atm CF$_4$ is estimated to be ~8 cm$^2$ at around 1MeV, which is larger than that of a typical non-imaging Ge semiconductor with 2"(diameter) ×2"(depth) (Fig.7b). Taking into account the effective background rejection via imaging, the ETCC can achieve ~10$^2$ times better sensitivity for line gammas than Ge detectors.

**Discussion**

Our success in gamma imaging-spectroscopy marks the establishment of proper gamma imaging similar to the normal optics, which enables us to measure the spectrum at any place, keeping information of the intensity or brightness. At present in 2016, technology of quantitative gamma imaging is highly coveted for effective decontamination in the area in and around Fukushima, Japan after the major accident of the nuclear power plant in 2011. Although many CCs have been developed and deployed there for the task, none has detected any more than some hot spots, and certainly none has reported a quantitative radiation map. Our interpretation of CCs explains the underlying reasons of those poor performances. In contrast, we have successfully obtained the first spectroscopic images in contamination areas, using an ETCC, which is reported in a separate paper[22].

The core technology for the ETCC has been tested in operations under intense radiation[13] and is now established. Our present model of the ETCC[13] has proved its superior characteristics. Its potential is very promising with the characteristics including a compatible effective area to Ge detectors, a wide FoV of several sr, a wide energy band from 50 to 10,000 keV, and a sharp PSF with the size of a few degrees. Another advantage of the ETCCs in practical uses is no need of expensive crystal or semiconductor, because a better SPD is more crucial than the better ARM to obtain a sharp PSF. The key technology for a sharp PSF (a few degree) is the way to measure the direction of recoil electrons in gas (see Fig.5 in TT). Although a SPD to achieve it requires very fine 3D-tracking of electrons at the scale of sub-μm in semiconductors, tracking with that precision has been already realized with a mm-size pixel readout of micro-pattern gas devices.

While nuclear gammas have been actively studied and utilized in the human society since the early 20$^{th}$C, the difficulty in proper imaging of gammas has been a major cause of stagnation in the field for decades. Now we expect that this advent of a method of clear visualization of nuclear gammas can, and likely will, open a new era to advance useful applications of radiation, as well as to promote the perceived safety of the radiation-related affairs.

**Methods**

**Reconstruction of the gamma-ray direction on the laboratory coordinates**

The ETCC enables us to reconstruct the direction of incident gamma-rays, based on the energies and directions of a Compton-scattered gamma and a recoil electron. Figures 1b and 1c shows the diagram of Compton-scattering in the laboratory coordinates. Let $\vec{s}, \vec{g}, \vec{e}$ and $\overrightarrow{e_{obs}}$ be the unit vectors of the direction of an incident gamma-ray, the direction of the scattered gamma-ray, the real direction of recoil electron and the observed direction of recoil electron, respectively. Let $K_e$ and $E_\gamma$ be the energy of recoil electron and scattered gamma-ray, respectively. In the laboratory coordinates, an incident gamma-ray has two angles: an angle $\zeta$ between z axis and $\vec{s}$, and the other η between X-axis and the direction of projected $\vec{s}$ to the X-Y plane. These angles are invariable, and satisfy the following equations:

$$\cos \zeta = -\vec{s} \cdot \overrightarrow{e_z} \tag{1}$$

$$\cos \eta = \frac{\overrightarrow{e_z} \times (\overrightarrow{e_z} \times \vec{s})}{|\overrightarrow{e_z} \times (\overrightarrow{e_z} \times \vec{s})|} \tag{2}$$

We define another two angles θ and ϕ to reconstruct the incident gamma-ray as shown in Fig.1b; they are the Compton-scattering angle and electron-azimuthal angle (between vectors projected $\vec{e}$ and $\overrightarrow{e_{obs}}$ to a plane perpendicular to $\vec{g}$), respectively, on the Compton coordinates. They are variable event by event, and are given as

$$\cos \theta = 1 - \frac{m_e c^2}{E_\gamma + K_e} \frac{K_e}{E_\gamma} \tag{3}$$

$$\cos \phi = \frac{\vec{g} \times (\vec{e} \times \vec{g})}{|\vec{g} \times (\vec{e} \times \vec{g})|} \cdot \frac{\vec{g} \times (\overrightarrow{e_{obs}} \times \vec{g})}{|\vec{g} \times (\overrightarrow{e_{obs}} \times \vec{g})|} \tag{4}$$

In this formulation, the ARM and SPD are the uncertainties of θ and ϕ, respectively.

The direction of the reconstructed incident gamma-ray $\vec{r}$ is described on the laboratory coordinates as follows:

$$\vec{r} = \frac{E_\gamma}{E_\gamma + K_e} \vec{g} + \frac{\sqrt{K_e(K_e + 2m_e c^2)}}{E_\gamma + K_e} \vec{e} \tag{5}$$

When the reconstruction is done correctly for the incident gammas from a source, the vector $\vec{r}$ corresponds to $\vec{s}$.

**Back projection imaging for nearby sources**

If we measure gammas from a source placed near the ETCC, the distribution of travelling direction of incident gammas detected by the ETCC are widely spread. In this case, back-projection imaging is more convenient than the image of the angular distribution of reconstructed travelling direction of gammas used in astronomy. We define a back-projection plane as illustrated in Fig.8, which includes the point of the source position. The intersection points of the infinite straight lines on reconstructed vectors calculated with the ETCC analysis with the back-projection plane are concentrated near the source position. The signal-to-noise ratio is improved by selecting events whose intersection points are around the source position like the area inside the dotted black line in Fig. 8. On the other hand, some circles of background gammas by the CC analysis fall on the source region. We can suppress the contamination of background events from the other region by the ETCC analysis.

**Further improvements of the ETCC**

The PSF and effective area achieved by the current ETCC are still insufficient for most of practical applications of nuclear gammas. In particular, a sharp drop of the effective area above 500 keV with the present ETCC (Fig.7b) is highly undesirable. The major cause is escape of the high-energy recoil electrons from the TPC; then, if side PSAs are installed inside the gas vessel, as opposed to outside as in the current ETCC (SMILE-II), they will catch these high-energy recoil electrons, and the performance should be greatly improved. Fig.7a illustrates the concept. Figure 7b shows the simulated result of the energy dependence of the effective areas for a 30-cm-cubic ETCC with 3 atm $CF_4$ gas or 2 atm Ar and GSO PSAs with thicker layers (26 mm for bottom PSAs and 19 mm for side ones). Their effective areas are calculated to be 8 cm$^2$ and 3 cm$^2$ at 1 MeV for 3-atm $CF_4$ and 2-atm Ar gases, respectively, which correspond to about 0.8% and 0.3% in gamma-ray detection efficiency, respectively. Since the ETCC fully reconstructs the Compton scattering process, its effective area (or detection efficiency) corresponds to the so-called photo-peak effective area (or efficiency) for general spectrometers, and those effective areas are roughly equivalent to 2-inch (diameter) x 2-inch (depth) Ge detectors (Fig.7b). A system with this design concept is now under construction, and will be completed in 2017.

We note that Gros, P. *et al*. 2016[23] has recently reported an elegant method to resolve the above-mentioned uncertainty in tracking in the X and Y strips readout system; they used a pulse height on each strip, searching for the correlation in the variation of pulse heights along the X and Y strips. Their method is applicable to our ETCC system by using the TOT information. We are trying to implement their analysis method, and are expecting that a good PSF of a few degrees will be achieved in near future with the ETCC without the use of complex pixel readout in the MPGD.

**References**

1. Todd, R.W., Nightingale, J. M. & Everett, D. B. A proposed γ camera. *Nature*. **251**, 132-134; 10.1038/251132a0 (1974).

2. Watanabe, S. *et al*. The Si/CdTe semiconductor Compton camera of the ASTRO-H Soft Gamma-ray Detector (SGD). *Nucl. Instrum. Meth. A*. **765**, 192-201; 10.1016/j.nima.2014.05.127 (2014).

3. Kierans, C. A. *et al*. Calibration of the Compton Spectrometer and Imager in preparation for the 2014 balloon campaign. *Proc. of SPIE*. **9144**, 91443M; 10.1117/12.2055250 (2014).

4. McCleskey, M. *et al*. Evaluation of a multistage CdZnTe Compton camera for prompt γ imaging for proton therapy. *Nucl. Instrum. Meth. A*. **785**, 163-169; 10.1016/j.nima.2015.02.030 (2015).

5. Vetter, K. Multi-sensor radiation detection, imaging, and fusion. *Nucl. Instrum. Meth. A*. **805**, 127-134; 10.1016/j.nima.2015.08.078 (2016).

6. Bloser, P. F., Legere, J. S., Bancroft, C. M., Ryan, J. M. & McConnell, M. L. Balloon flight test of a Compton telescope based on scintillators with silicon photomultiplier readouts. *Nucl. Instrum. Meth. A*. **812**, 92-103; 10.1016/j.nima.2015.12.047 (2016).

7. Atwood, W. B. *et al*. The Large Area Telescope on the Fermi Gamma-ray Space Telescope Mission. *Astrophys. J*. **697**, 1071-1102; 10.1088/0004-637X/697/2/1071 (2009).

8. Schönfelder, V. *et al*. Instrument description and performance of the Imaging Gamma-Ray Telescope COMPTEL aboard the Compton Gamma-Ray Observatory. *Astrophys. J. Suppli*. **86**, 657-692; 10.1086/191794 (1993).

9. Varendoff, M. G. *et al*. COMPTEL images locations of gamma-ray bursts. *AIP conf. Proc*. **265**, 77-81; 10.1063/1.42782 (1991).

10. Bloemem, H. *et al*. COMPTEL observations of the Orion complex: evidence for cosmic-ray induced gamma-ray line. *Astron. Astrophys*. **281**, L5-L8 (1994).

11. Strong, A. W. *et al*. Diffuse Galactic Continuum Emission: Recent Studies using COMPTEL Data. *AIP Conf. Proc*. **410**, 1198-1202; 10.1063/1.53932 (1997).

12. Diehl, R. *et al*. 26Al imaging details from COMTEL. *Adv. Space Res*. **15**, 123-126; 10.1016/0273-1177(94)00050-B (1995).

13. Tanimori, T. *et al*. An Electron-Tracking Compton Telescope for a Survey of the Deep Universe by MeV Gamma-Rays. *Astrophys. J*. **810**, 28; 10.1088/0004-637X/810/1/28 (2015).

14. Tümer, O. T. *et al*. The TIGRE Instrument for 0.3-100 MeV Gamma-Ray Astronomy. *Transaction on Nuclear Science IEEE*. **42**, 907-916; 10.1109/23.467770 (1995).

15. Kanbach, G. *et al*. MEGA-A Next Generation Mission in Medium Energy Gamma-Ray Astronomy. *AIP Conf. Proc*. **587**, 887-891; 10.1063/1.1419516 (2001).

16. Zoglauer, A., Andritschke, R., Bloser, P. & Kanbach, G. Imaging Properties of the MEGA Prototype. *Nuclear Science Symposium Conference Records 2003 IEEE*. **3**, 1694-1698; 10.1109/NSSMIC.2003.1352205 (2004).

17. Takada, A. *et al*. Observation of Diffuse Cosmic and Atmospheric Gamma Rays at Balloon Altitudes with an Electron-Tracking Compton Camera. *Astrophys. J*. **733**, 13; 10.1088/0004-637X/733/1/13 (2011).

18. Mizumoto, T. *et al*. New readout and data-acquisition system in an electron-tracking Compton camera for MeV gamma-ray astronomy (SMILE-II). *Nucl. Instrum. Meth. A*. **800**, 40-50; 10.1016/j.nima.2015.08.004 (2015).

19. Agostinelli, S. *et al*. Geant4—a simulation toolkit. *Nucl. Instrum. Meth. A*. **506**, 250-303; 10.1016/S0168-9002(03)01368-8 (2003).

20. Mizumoto, T. *et al*. A performance study of an electron-tracking Compton camera with a compact system for environmental gamma-ray observation. *J. of Instrum*. **10**, C06003; 10.1088/1748-0221/10/06/C06003 (2015).

21. Bandstra, M. S. *et al*. Detection and Imaging of the Crab Nebula with the Nuclear Compton Telescope. *Astrophys. J*. **738**, 8; 10.1088/0004-637X/738/1/8 (2011).

22. Tomono, D. et al. First On-Site True Gamma-Ray Imaging-Spectroscopy of Contamination near Fukushima Plant. Sci. Rep. (SREP-16-38035B).

23. Gros, P. *et al*. First measurement of polarisation asymmetry of a gamma-ray beam between 1.74 to 74 MeV with the HARPO TPC. *Proc. of SPIE*, **9905**, 99052R; 10.1117/12.2231856 (2016).
**Acknowledgements**

This study was supported by the Japan Society for the Promotion of Science (JSPS) Grant-in-Aid for Scientific Research (S) (21224005), JSPS Grant-in-Aid for Challenging Exploratory Research (25610042), Grant-in-Aid for young scientists (B) (15K17608), a Grant-in-Aid from the Global COE program "Next Generation Physics, Spun from Universality and Emergence" from the Ministry of Education, Culture, Sports, Science and Technology (MEXT) of Japan. Some of the authors thank for Grant-in-Aid for JSPS Fellows (11J00606, 13J01213, 16J08498).
**Author contributions statement**

T.Tan., is a leader of this experiment and wrote the manuscript.

Y.Mi. mainly led the measurements of imaging spectroscopy and wrote the manuscript/

T.Tak is a sub leader and lead the simulation and wrote the manuscript.,

S.M. contributed to develop the analysis method for imaging spectroscopy.

T.Tak contributed to the simulation results and made figures.

T.K contributed to operate the ETCC and measure the spectroscopic data/.

S.Ko contributed to construct the ETCC used here and make analysis software for noise reduction.

H.K.contributed to develop the PSAs in particular GSO crystals.

S.Ku. contributed to develop the PSAs in particular readout electronics.

Y.Ma. contributed to develop the data acquisition system.

K, M. contributed to gas vessel and gas system.

T,M. contributed to develop the readout system of the TPC and its reconstruction software.

Y.N. contributed to develop the photo sensor of PSAs.

K.N. contributed to the gas study of TPC.

J.D.P. contributed to develop the MPGD and power unit.

T.S. contributed to construct the ETCC and simulation program.

S.S. contributed to support the measurement of imaging spectroscopy.

D.T. contributed to the improvement of electron tracking analysis.

K.Y. contributed to develop the MPGD

**Competing financial interests**

The authors declare no competing financial interests.

**Corresponding Author**

Correspondence and requests for materials should be addressed to T.T. (tanimori@cr.scphys.kyoto-u.ac.jp)

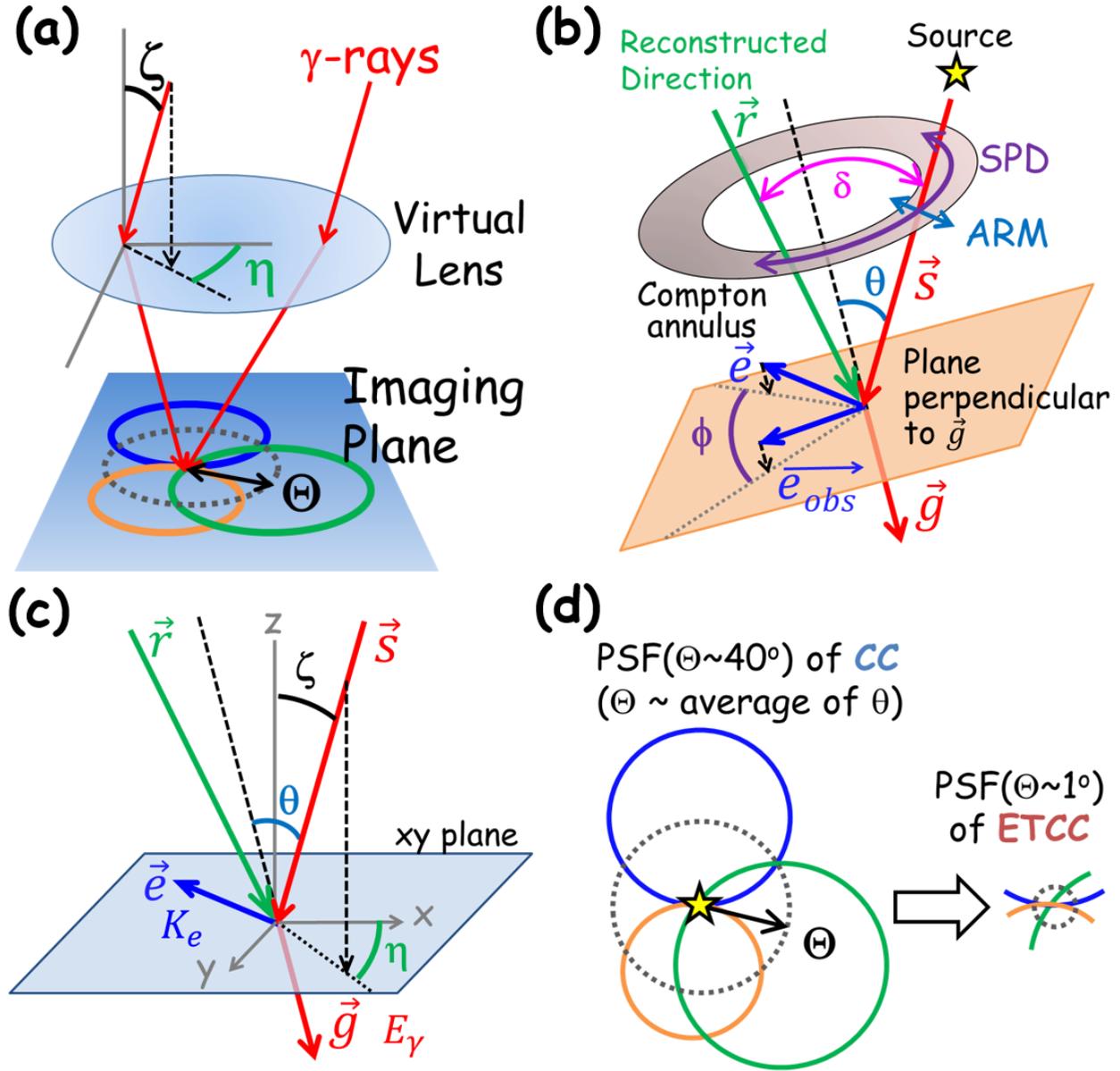

**Figure 1. Schematic explanations of gamma-ray reconstruction, the PSF and coordinates systems.** (a) Schematic diagram of the PSFs of the CC and ETCC on the imaging plane (laboratory coordinates). An image of a point source is displayed with event circles reconstructed in CC. Here ζ and η denote two angles of an incident gamma. (b) Schematic explanation of the Compton scattering kinematics. The parameters θ and ϕ denote the Compton-scattering angle and electron-azimuthal angles of the incident gamma, respectively, on the Compton coordinates. (c) Coordinate diagram for explanation of the angles of ζ, η, and θ in the laboratory coordinates x-y-z. (d) Schematic explanation of PSF(Θ) of the CC and ETCC. See text for the definition of Θ.

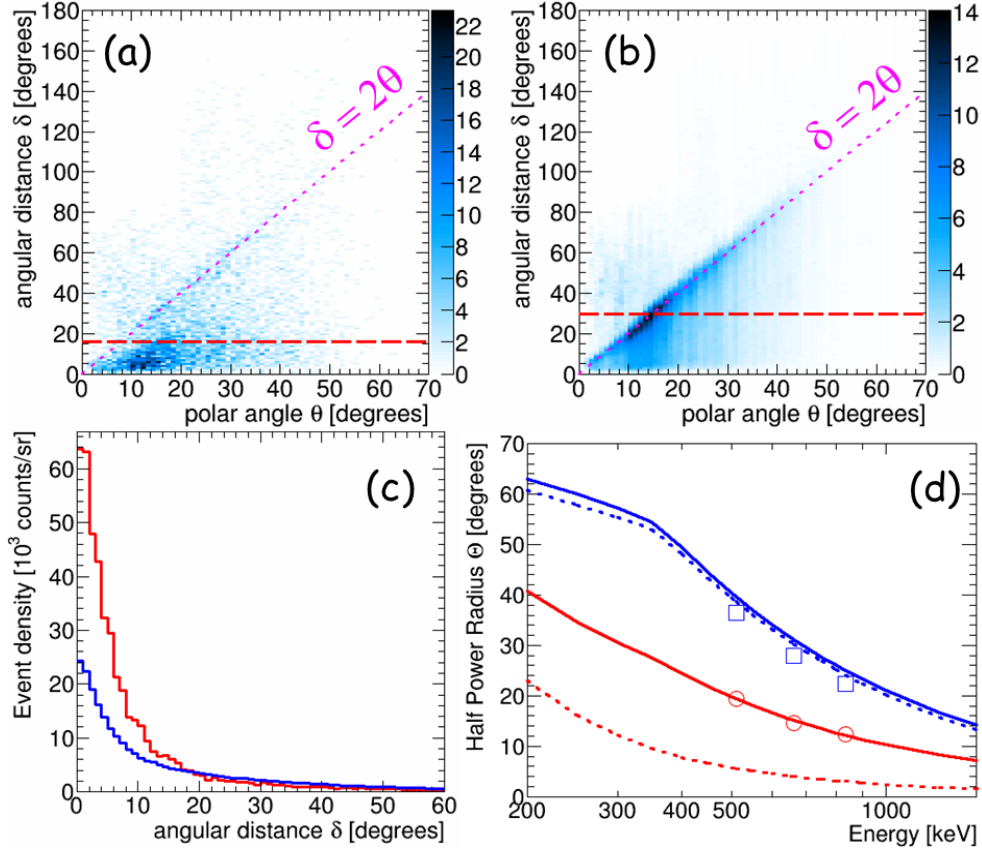

**Figure 2. Measured and simulated correlations of δ and θ in CC and ETCC analyses, and their PSFs.** (a, b) Measured scatter plots of the angular distance between the reconstruction direction and the real direction δ versus the Compton-scattering angle θ derived with the ETCC and CC analyses, respectively. Panel (a) is a scatter plot of δ determined by θ and ϕ in the ETCC analysis, whereas panel (b) is that of the event-density distribution obtained in the CC analysis. Red dashed-lines indicate the half power radius Θ of these PSFs. (c) Measured event-density distributions per unit solid angle as a function of δ for the ETCC analysis (red) and CC analysis (blue). In this figure, the total number of events with both the ETCC and CC analyses are the same, each of which is calculated by integrating the event density at δ multiplied by $2\pi\delta$ for the range of δ up to $2\pi$. (d) Measured variation of Θ with the (blue squares) CC and (red circles) ETCC analyses as a function of gamma-ray energy. Red solid and dotted lines show the simulation results with the present and improved (ARM=2° and SPD=15° at 662 keV) ETCCs, respectively, where the energy dependences of ARM and SPD are taken into account. Blue solid and dotted lines are the same simulation results of the CC analysis with ARM=5° (present) and 2° (improved), respectively.

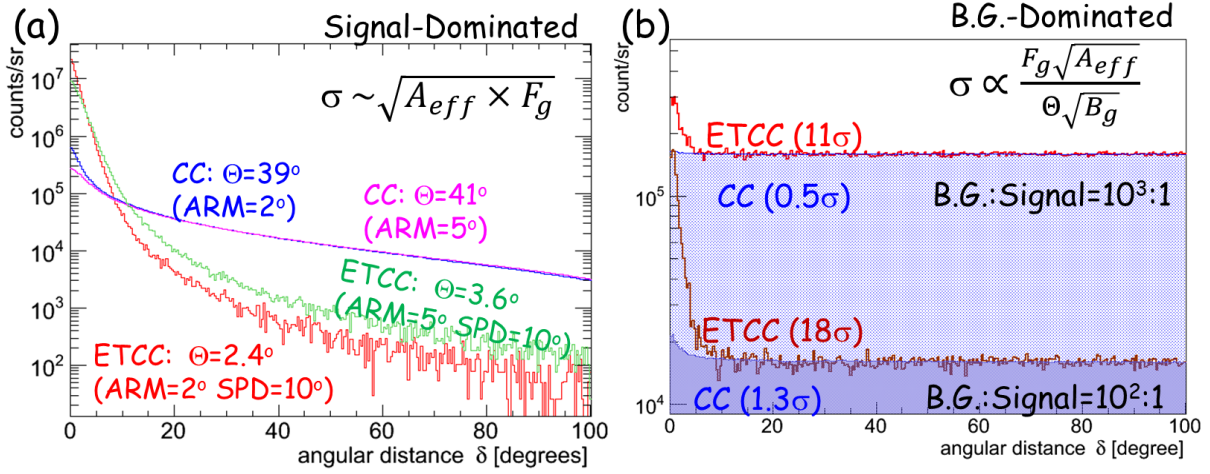

**Figure 3. Comparisons of simulated PSFs between CC and ETCC analyses.** (a) Simulated distributions of the event density per unit solid angle as a function of the angular distance δ for a point source in a signal-dominated case with the CC and ETCC analyses with the improved 30-cm-cubic ETCC (ARM=2º or 5º, and SPD=10º at 662 keV) (b) Same as (a) but in background-dominated cases with the background-to-signal ratios of $10^2$ and $10^3$.

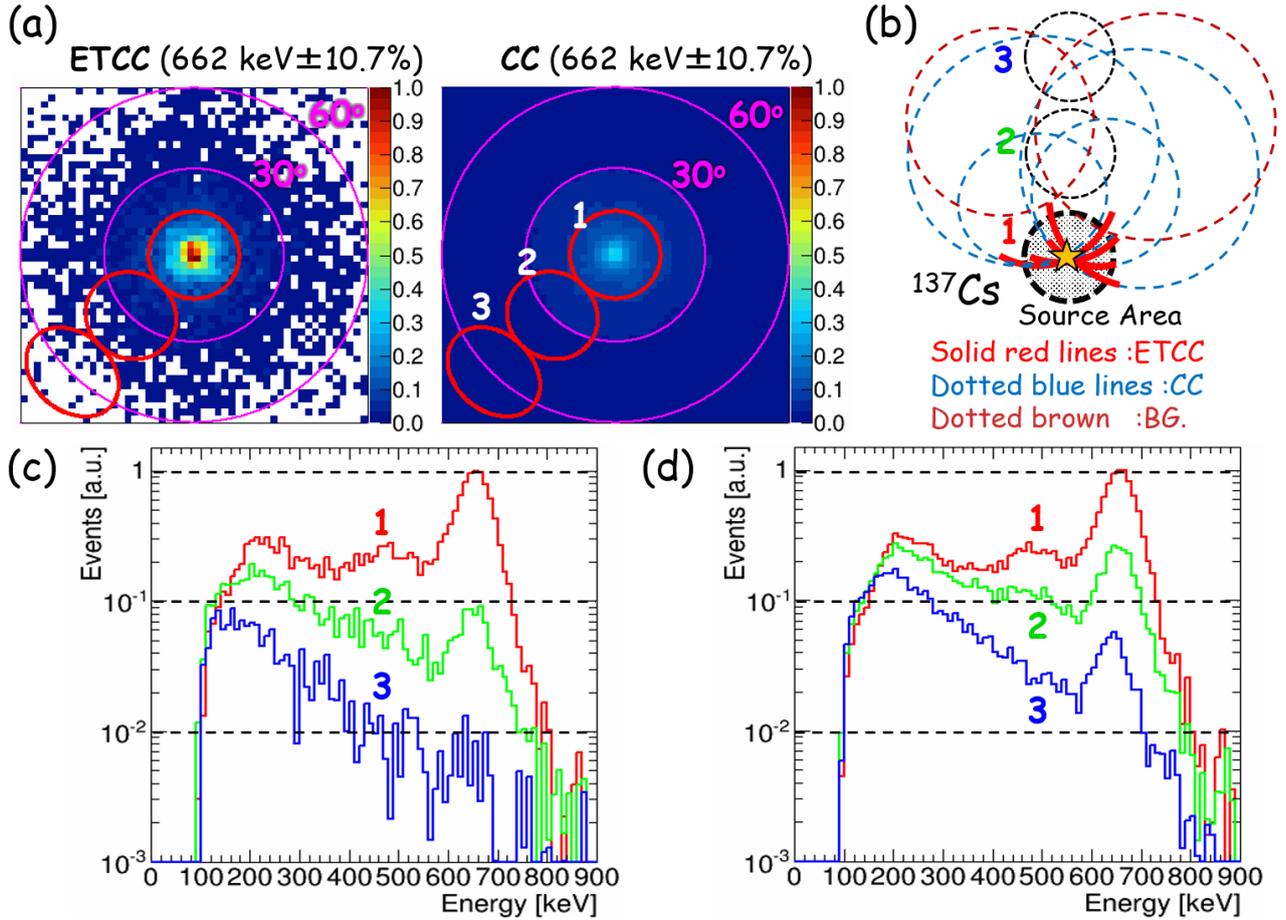

**Figure 4. Comparison of measured images and event leakages in CC and ETCC analyses.** (a) Measured images of 662 keV point source ($^{137}$Cs: 0.86 MBq) placed at a distance of 2 m with the ETCC analysis (left panel) and CC analysis (right). The colour-bars in both the panels indicate the ratio of the event density to the maximum value in the ETCC analysis. The red circles (labelled the numbers of 1, 2 and 3) are 15°-radius regions for the spectra indicated in panels (c) and (d). (b) A schematic drawing to illustrate the difference of the outward leakage of events in the PSF between the ETCC and CC analyses. (c and d) Energy spectra calculated with the (c) ETCC and (d) CC analyses, after the gammas for each spectrum are accumulated from a circular region with a radius of 15°, located at the angular distances of 0° (red, labelled 1), 30° (green, labelled 2), and 60° (blue, labelled 3) from the centre of each region to the source position. In each spectrum, the measured events are normalized to the maximum number of events at 662 keV at the centre of FoV.

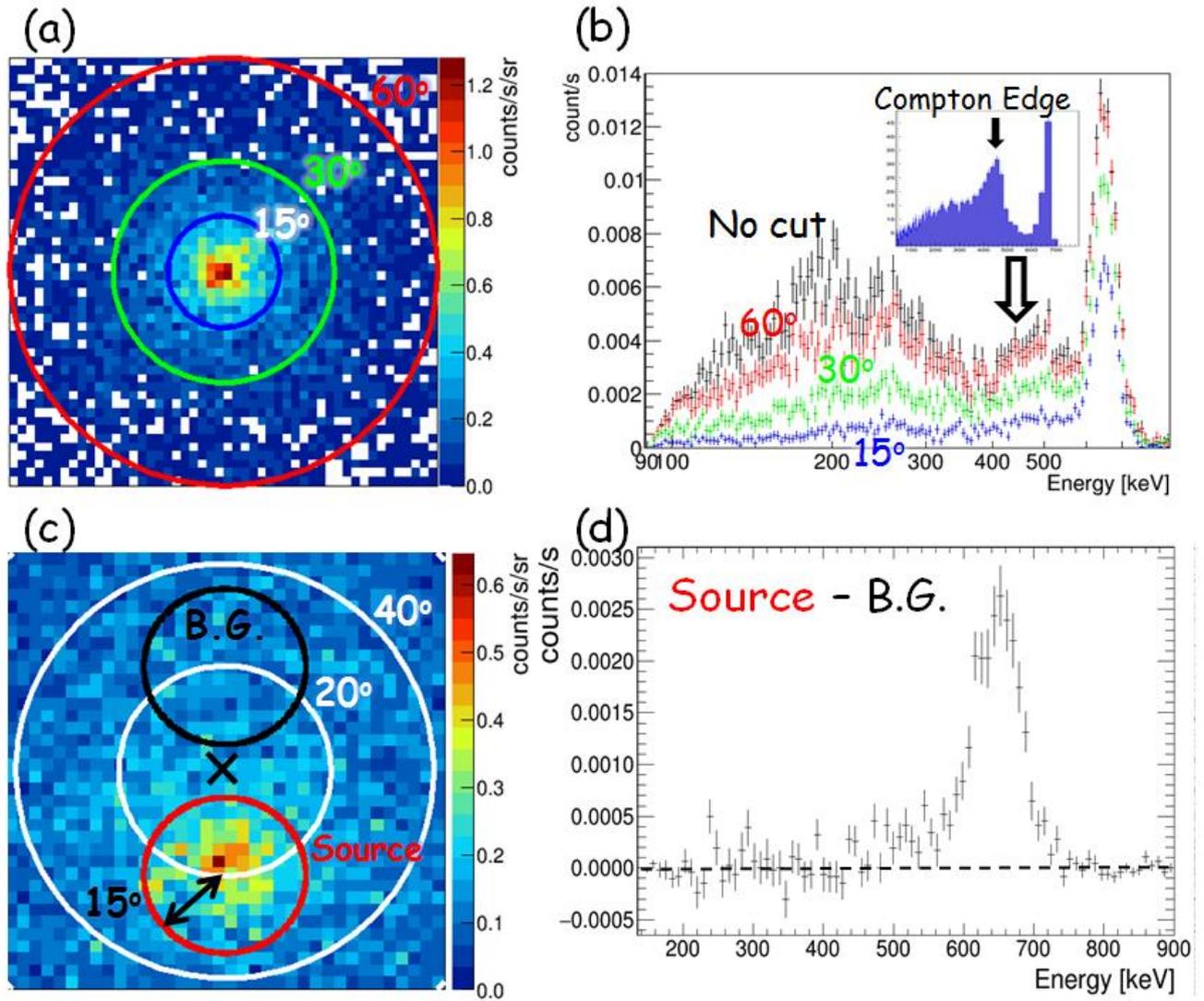

**Figure 5. Results of imaging spectroscopy using a $^{137}$Cs radioisotope.** (a, b) An image for the whole energy band and energy spectra of gammas from $^{137}$Cs (0.86 MBq) source placed at a distance of 2 m from the ETCC with angular radii of 15º (blue), 30º (green), 60º (red), and 180º (black) between the extracted region and the source. A simulated energy spectrum is, for comparison, given in the inset in Panel (b) for events with recoil electrons hitting the backward detector. (c) A image of $^{137}$Cs (2.9 MBq) source at the off-axis with a polar angle of 20º. Red and black circles, each of which is 15º in radius, indicate source and background regions, respectively. (d) The spectrum after air-scattered gammas within the PSF(Θ~15º) are removed, using the region symmetrical about the centre of the FoV (black circle in c) as the background region.

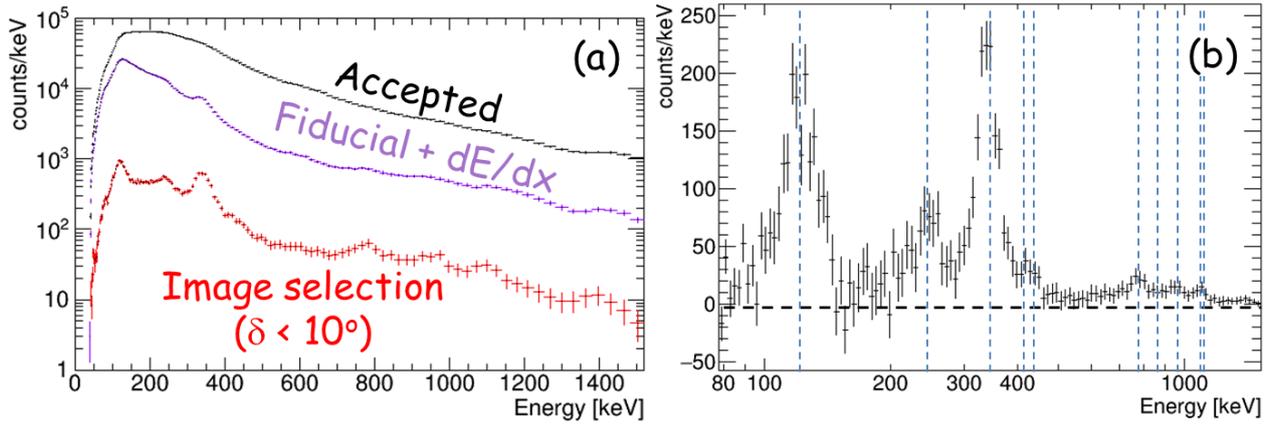

**Figure 6. Demonstration of background rejection by imaging spectroscopy using a $^{152}$Eu radioisotope.** (a) Energy spectra of gammas of $^{152}$Eu (0.72 MBq) placed off-axis at a polar angle of 30° with a distance of 2.4 m. Black, purple, and red crosses are the energy spectra for all the events, reconstructed gamma-ray events (see text for the filtering criteria), and the events after the imaging selection ($\delta < 10°$), respectively. (b) Energy spectra with the remaining background subtracted, which is accumulated from a symmetrical region indicated in Fig. 5d. Vertical blue dashed lines are major gamma-line energies of $^{152}$Eu between 100 and 1200 keV.

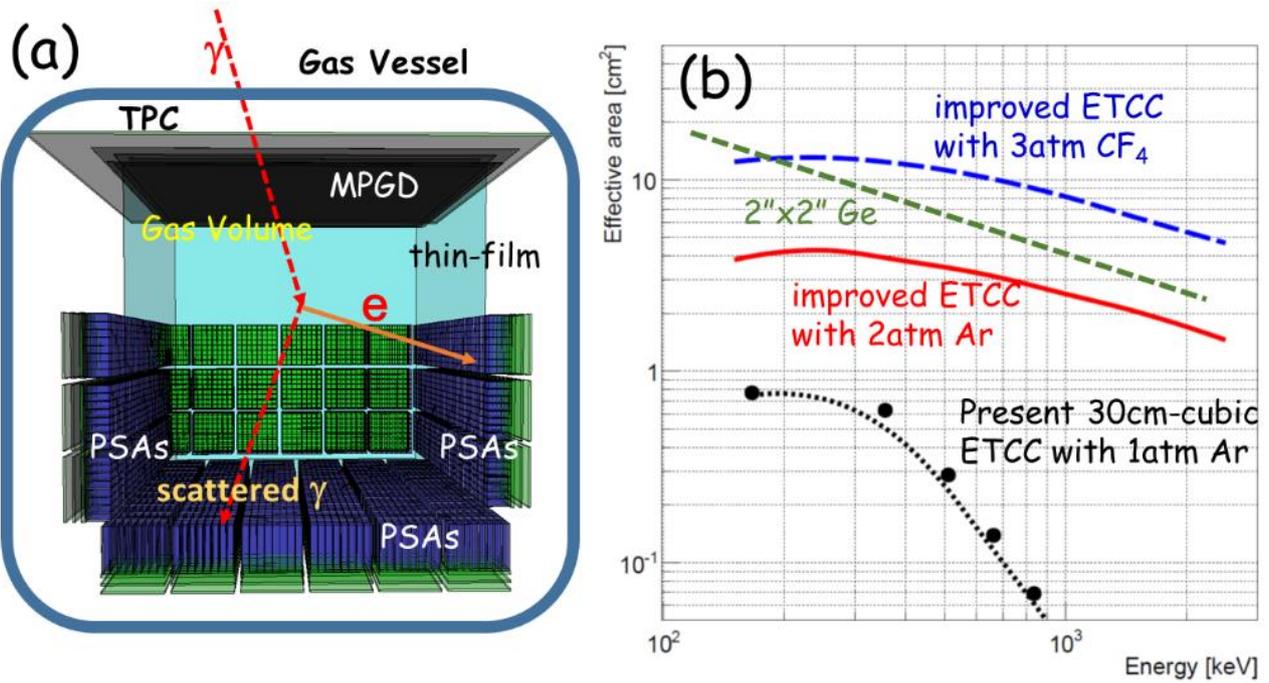

**Figure 7. Schematic view and effective areas of future-planed ETCC.** (a) Schematic view of the inside of the TPC vessel of the revised ETCC. (b) Energy dependence of the effective areas. The filled-circles are the measured effective area of the present 30-cm-cubic ETCC with Ar 1-atm gas (SMILE-II ETCC), and the black dotted line is its simulated results. The red-solid and blue-dashed lines indicate the simulated effective areas for 30-cm-cubic ETCCs with 2-atm Ar gas and with 3-atm $CF_4$ gas, respectively, with the PSA installed inside. Green dashed line is the effective area for the line gammas of 2" (diameter) x 2" (depth) Ge detectors.

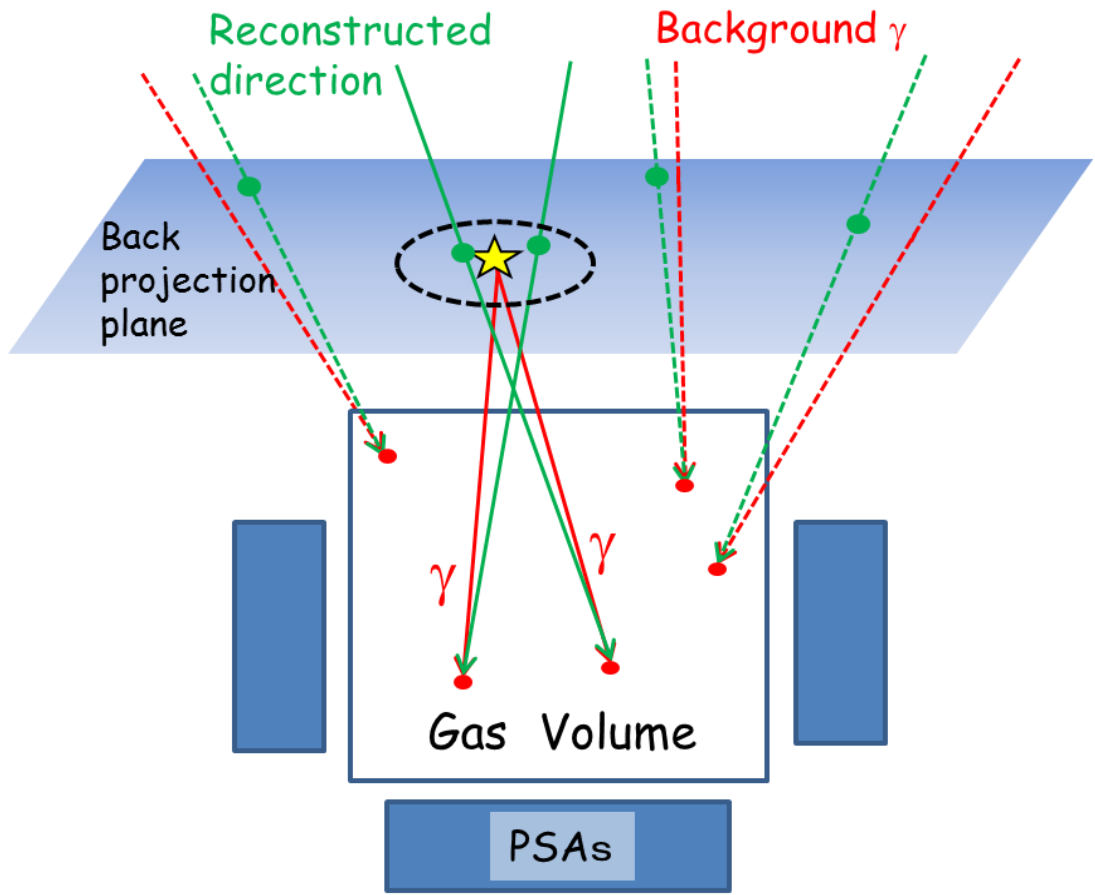

**Figure 8. Schematic explanation of back projection imaging.** Schematic view of back-projection imaging. The red and green arrows show the vectors of actual incident gammas and reconstructed directions with the ETCC analysis, respectively. The back-projection plane encompasses the point of the source position (the yellow star). The red and green points show the Compton scattering points and intersection points, respectively, of reconstructed gamma-ray tracks on the back-projection plane.